%% file: Paper_13.10.13_4_2.tex
\documentclass[letterpaper,11pt]{article}

\usepackage{amsmath,amssymb,epsfig,bbm}
\usepackage[active]{srcltx}
\usepackage[all]{xypic}

\usepackage{color}

\usepackage{dsfont}

\definecolor{co}{cmyk}{0,0.7,0.3,0}
\definecolor{darkgreen}{cmyk}{1,0,1,.2}
\definecolor{m}{rgb}{1,0.1,1}

\newcommand{\be}{\begin{equation}}
\newcommand{\ba}{\begin{eqnarray}}
\newcommand{\ea}{\end{eqnarray}}
\newcommand{\nn}{\nonumber}

\def\d{\delta}
\def\e{\epsilon}

\def\oo{\omega}
\def\p{\pi}

\def\G{\Gamma}

\def\P{\Pi}

\def\ca{{\cal A}}
\def\cb{{\cal B}}

\def\cd{{\cal D}}

\def\cf{{\cal F}}

\def\ch{{\cal H}}

\def\cl{{\cal L}}

\def\co{{\cal O}}

\newcommand{\C}{{\Bbb C}}

\newtheorem{thm}{Theorem}[subsection]

\newtheorem{definition}[thm]{Definition}

\fontfamily{yfrak}

\begin{document}

\vskip 25mm

\begin{center}

{\Large\bfseries The Quantum Holonomy-Diffeomorphism Algebra \&  Quantum Gravity
}

\vskip 4ex

Johannes \textsc{Aastrup}$\,^{a}$\footnote{email: \texttt{aastrup@math.uni-hannover.de}} \&
Jesper M\o ller \textsc{Grimstrup}\,\footnote{email: \texttt{jesper.grimstrup@gmail.com}}\\ 
\vskip 3ex  

$^{a}\,$\textit{Mathematisches Institut, Universit\"at Hannover, \\ Welfengarten 1, 
D-30167 Hannover, Germany.}

\end{center}

\vskip 3ex

\begin{abstract}

We introduce the Quantum Holonomy-Diffeomorphism $*$-algebra, which is generated by holonomy-diffeomorphisms on a 3-dimensional manifold and translations on a space of $SU(2)$-connections. We show that this algebra encodes the canonical commutation relations of canonical quantum gravity formulated in terms of Ashtekar variables. Furthermore, we show that semi-classical states exist on the holonomy-diffeomorphism part of the algebra but that these states cannot be extended to the full algebra. Via a Dirac type operator we derive a certain class of unbounded operators that act in the GNS construction of the semi-classical states. These unbounded operators are the type of operators, which we have previously shown to entail the spatial 3-dimensional Dirac operator and Dirac Hamiltonian in a semi-classical limit.
Finally, we show that the structure of the Hamilton constraint emerges from a Yang-Mills type operator over the space of $SU(2)$-connections.

\end{abstract}

\newpage
\tableofcontents

\section{Introduction}

In the quest for a quantum theory of gravity the choice of an algebra, which in one way or another involves the quantum variables of general relativity, is a pivotal first step. In the present situation, where direct experimental verification is beyond reach, it is our belief that one must search for an algebra with a high degree of canonicity and naturalness. \\

In this paper we present a $*$-algebra, which displays a high degree of both canonicity and naturalness. The algebra, which can be understood as a quantum mechanics on a space of connections, involves two sets of variables: a choice of functions over the space of connections and translations hereon. The functions are the holonomy-diffeomorphisms of a 3-dimensional manifold -- encoding how objects are moved in space -- and the interaction between these holonomy-diffeomorphisms and the translation operators encodes the quantized commutation relations of general relativity formulated in terms of Ashtekar variables \cite{Ashtekar:1986yd,Ashtekar:1987gu}. \\

We call this algebra the Quantum Holonomy-Diffeomorphism ($\mathbf{QHD}$) algebra and a theory based on this algebra for Quantum Holonomy Theory. \\

In terms of canonicity, the $\mathbf{QHD}$ algebra is canonical up to the dimension of the manifold $M$, which we set equal to three, and up to a choice of gauge group, where our choice of $SU(2)$ partly follows from the dimension of $M$. The translation operators are canonical. 

In terms of naturalness, we believe that an algebra generated by holonomy-diffeomorphisms is the single most natural object to consider when attempting to quantize gravity, since this algebra encodes how diffeomorphisms act on spinors. 
Furthermore, adding a 'quantum' to the algebra of holonomy-diffeomorphisms, and thereby constructing a quantum mechanics on a space of connections, is not only canonical but also an exceedingly natural thing to do. Such a construction automatically includes the commutation relations of canonical quantum gravity.\\

With this algebra at hand a natural next step is to search for states and to apply the GNS-construction in order to obtain a Hilbert space representation. Using lattice approximations we show that a state exist on the first part of the algebra -- the holonomy-diffeomorphisms -- but that such a state cannot be extended to the full $\mathbf{QHD}$-algebra. It is therefore not clear whether states on $\mathbf{QHD}$ exist. We find, however, that there exist another class of unbounded operators, which act in the GNS construction of this state. We find these additional operators via an auxiliary Dirac type operator: they emerge from the second commutator of the Dirac type operator with elements in the holonomy-diffeomorphism algebra. It turns out that these operators are the type of operators, which we have previously shown to entail the spatial 3-dimensional Dirac operator and Dirac Hamiltonian in a semi-classical analysis, see \cite{Aastrup:2012jj}-\cite{AGNP1} (see also  \cite{AGN3}-\cite{Aastrup:2005yk} for earlier versions).  \\

Furthermore, we find that the states on the holonomy-diffeomorphism algebra will always be semi-classical. This means that each semi-classical approximation provides a different kinematical Hilbert space and that no other type of kinematical Hilbert space appear to exist.  \\

Finally, with the auxiliary Dirac type operator, we can write down one-forms and curvature operators. A simply computation shows that the structure of the Hamiltonian for General Relativity - formulated in terms of Ashtekar variables - emerges from the square of such a curvature operator. \\

This paper is organized as follows: In section 2 we first introduce the Holonomy-Diffeomorphism ($\mathbf{HD}$) algebra, which was first described in \cite{Aastrup:2012vq,AGnew}. We then introduce translations on the space of $SU(2)$ connections and define the Quantum Holonomy-Diffeomorphism algebra in section 2.2. In section 2.3 we show that this algebra encodes the canonical commutation relations of canonical quantum gravity. Section 3 is concerned with the possibility of construction semi-classical states on the $\mathbf{QHD}$ algebra via an infinite sequence of lattice approximations. Section 3.1 and 3.2 present the setup and in section 3.3 we confirm that a state exist on the $\mathbf{HD}$-part of the algebra and that this state cannot be extended to the full algebra. In section 4 we introduce an auxiliary construction with a Dirac type operator, which leads to a class of unbounded operators that act in the GNS construction of the semi-classical state. In section 5 we provide a simple computation that shows that the structure of the Hamiltonian of General Relativity emerges in a semi-classical limit from a curvature operator built from the auxiliary Dirac type operator. Finally, we conclude in section 6.

\section{The Quantum Holonomy-Diffeomorphism algebra}

The first task is to introduce the Holonomy-Diffeomorphism algebra. This algebra was first described in \cite{Aastrup:2012vq, AGnew}, where its spectrum was also analyzed.\\

\subsection{The Holonomy-Diffeomorphism algebra}

\begin{figure}[t]
\begin{center}
\resizebox{!}{4cm}{
 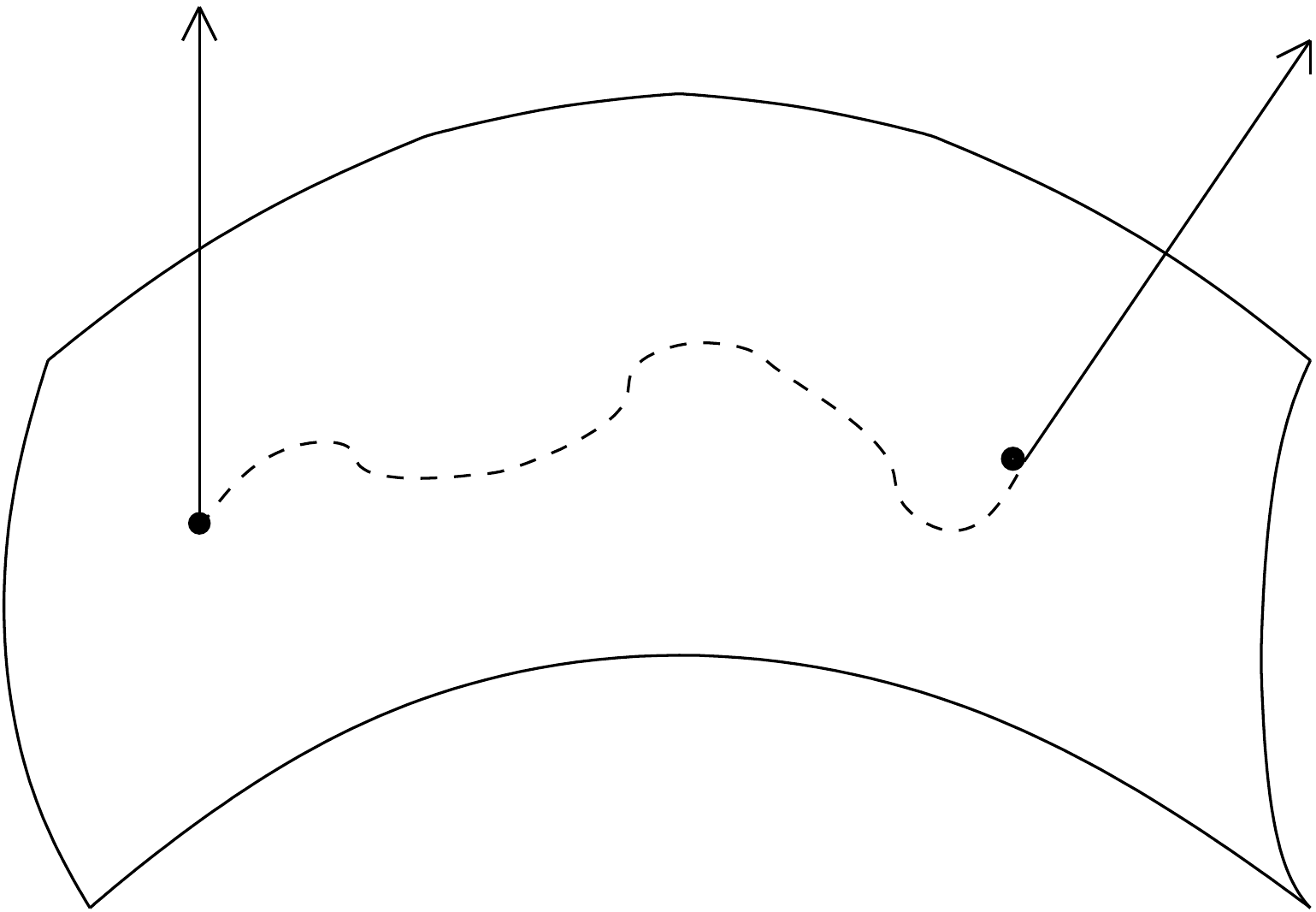}
\end{center}
\label{lenin}
\caption{An element in $\mathbf{H D}$ will parallel transport a vector on $M$ along the flow of a diffeomorphism.}
\end{figure}

Let $M$ be a connected $3$-dimensional manifold. We consider the two dimensional trivial vector bundle $S=M\times \C^2$ over $M$, and we consider the space of $SU(2)$ connections acting on the bundle. Given a metric $g$ on $M$ we get the Hilbert space $L^2(M,S,dg)$, where we equip $S$ with the standard inner produkt. Given a diffeomorphism $\phi:M\to M$ we get a unitary operator $\phi^*$ on  $L^2(M,S,dg)$ via
$$( \phi^* (\xi ))(\phi (m) )= (\Delta \phi )(m)  \xi (m) , $$
where  $\Delta \phi (m)$ is the volume of the volume element in $\phi (m)$ induced by a unit volume element in $ m$ under $\phi $.      

Let $X$ be a vectorfield on $M$, which can be exponentiated, and let $\nabla$ be a $SU(2)$-connection acting on $S$.  Denote by $t\to \exp_t(X)$ the corresponding flow. Given $m\in M$ let $\gamma$ be the curve  
$$\gamma (t)=\exp_{t} (X) (m) $$
running from $m$ to $\exp_1 (X)(m)$. We define the operator 
$$e^X_\nabla :L^2 (M , S, dg) \to L^2 (M ,  S , dg)$$
in the following way:
we consider an element $\xi \in L^2 (M ,  S, dg)$ as a $\C^2$-valued function, and define 
$$  (e^X_\nabla \xi )(\phi (m))=  ((\Delta \exp_1) (m))  \hbox{Hol}(\gamma, \nabla) \xi (m)   .$$
Here  $\hbox{Hol}(\gamma, \nabla)$ denotes the holonomy of $\nabla$ along $\gamma$. Again, the factor $(\Delta \exp_1) (m)$ is accounting for the change in volumes, rendering $e^X_\nabla$ unitary.  \\

Let $\ca$ be the space of $SU(2)$-connections. We have an operator valued function on $\ca$ defined via 
$$\ca \ni \nabla \to e^X_\nabla  . $$
We denote this function $e^X$. 

Denote by $\cf (\ca , \cb (L^2(M, S,dg) ))$ the bounded operator valued functions over $\ca$. This forms a $C^*$-algebra with the norm
$$\| \Psi \| =  \sup_{\nabla \in \ca} \{\|  \Psi (\nabla )\| \}, \quad \Psi \in  \cf (\ca , \cb (L^2(M, S,dg )) ) $$

For a function $f\in C^\infty_c (M)$ we get another operator valued function $fe^X$ on $\ca$.

\begin{definition}
Let 
$$C =   \hbox{span} \{ fe^X |f\in C^\infty_c(M), \ X \hbox{ exponentiable vectorfield }\}  . $$
The holonomy-diffeomorphism algebra $\mathbf{H D} (M,S,\ca)   $ is defined to be the $C^*$-subalgebra of  $\cf (\ca , \cb (L^2(M,S,dg )) )$ generated by $C$.

We will often denote $\mathbf{H D} (M,S,\ca)   $ by $\mathbf{H D}   $ when it is clear which $M$, $S$, $\ca$ is meant.
We will by $\ch \cd (M,S,\ca)   $ denote the  $*$-algebra generated by $C$.
\end{definition}

It was shown in \cite{AGnew} that the  $\mathbf{H D} (M,S,\ca)   $ is independent of the chosen metric $g$.

\subsection{The Quantum Holonomy-Diffeomorphism algebra}

Let $\mathfrak{su}(2)$ be the Lie-algebra of $SU(2)$. It is well-known that two connections in $\mathcal{A}$ differs by an element in $\Omega^1(M,\mathfrak{su}(2))$, and that for $\nabla \in \mathcal{A}$ and $\omega \in \Omega^1(M,\mathfrak{su}(2))$, $\nabla +\omega$ defines a connection in $\mathcal{A}$.  Thus  a section $\omega \in \Omega^1(M,\mathfrak{su}(2))$ induces a transformation of $\ca$, and therefore an operator $U_\omega $ on $\mathcal{F}(\ca,  \cb(L^2 (M ,  S,g)))$ via   
$$U_\omega (\xi )(\nabla) = \xi (\nabla - \omega) .$$ 

Note that $U^{-1}_\omega=U_{-\omega}$. 

\begin{definition}

Let us first denote by $\mathbf{QHD}(M,S,\ca)$ the sub-algebra of $\cf (\ca , \cb (L^2(M, S)) )$ generated by $\mathbf{HD}(M,S,\ca)$ and all the operators $U_{ \omega} $, $\omega \in \Omega^1(M,\mathfrak{su}(2))$. 
We will often denote $\mathbf{QHD}(M,S,\ca)$ by $\mathbf{QHD}$ when it is clear, which $M$, $S$, $\ca$ is meant. 
We call $\mathbf{QHD}$ the Quantum-Flow algebra or the Quantum Holononomy-Diffeomorphism algebra.
\end{definition}

We note that we have the relation 
\begin{equation} \label{konj}
(U_{\omega}f e^X U_{ \omega}^{-1}) (\nabla) =f e^X (\nabla + \omega )  , 
\end{equation}
where $f\in C^ \infty_c(M)$. However $\mathbf{QHD}$ is not a cross product of $\mathbf{HD}$ with the additive group $\Omega^1 (M,\mathfrak{su}(2))$, since the function of operators given by $\nabla \to e^X_{\nabla +  \omega}$ need not be in $\mathbf{HD}$.

\subsubsection{The infinitesimal $\mathbf{QHD}$ algebra}

To get closer to the formulation of the holonomy-flux-algebra\footnote{By the holonomy-flux algebra we refer to the algebra used in Loop Quantum Gravity, see \cite{AL1}.} and canonical quantization of gravity (see \cite{Aastrup:2012jj} for setup and notions) we need the infinitesimal version of $U_{t \omega}$. We simply do this by formally defining 
$$E_\omega  =\frac{d}{dt}U_{  t  \omega}|_{t=0} . $$
Due to the relation (\ref{konj}) we get 
\begin{equation} \label{flowkan}
[ E_\omega , e^X_\nabla ]= \frac{d}{dt}e^X_{\nabla +t\omega}|_{t=0}  . 
\end{equation}
Thus the infinitesimal version of the Quantum Holonomy-Diffeomorphism algebra is generated by the flows and the variables $\{ E_\omega\}_{\omega \in \Omega (M,T)}$. We denote this algebra by $\mathbf{dQHD}$. \\

We note, that 
$$
E_{\omega_1+\omega_2}=E_{\omega_1}+E_{\omega_2\;.}
$$
This follows, since the map $\Omega^1 (M,\mathfrak{su}(2))\ni \omega \to U_{ \omega}$ is a group homomorphism, i.e. $U_{(\omega_1+\omega_2 )}=U_{\omega_1}U_{ \omega_2}$. \\

\begin{figure}[t]
\begin{center}
\resizebox{!}{ 4 cm}{
 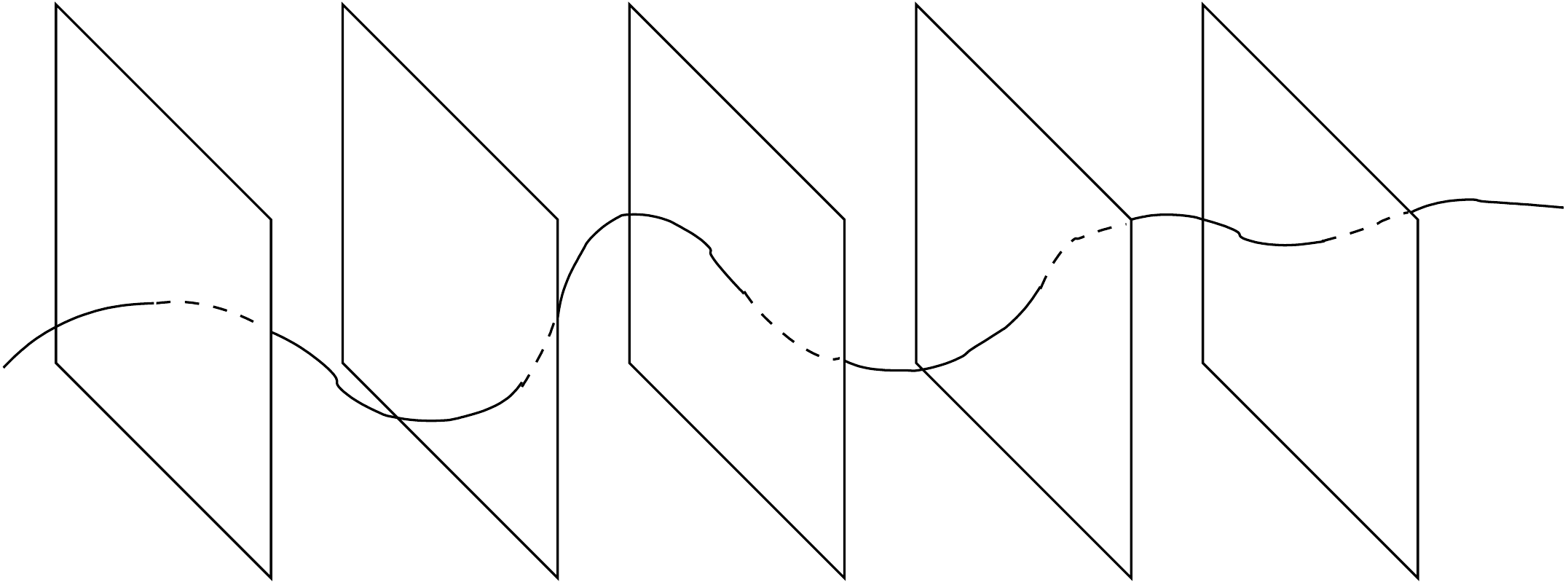}
\end{center}
\caption{\it The operator $E_{\sigma_i dx^j}$ will, when it is commuted with a flow, insert Pauli matrices continuously along the course of the flow. This means that it acts as a sum of flux operators with surfaces, which intersect the flow at the points of insertion. }
 \label{heltvildtsvedig}
\end{figure}

To see the connection to the holonomy-flux algebra let us analyze the righthand side of (\ref{flowkan}). First we introduce local coordinates $(x_1 ,x_2,x_3)$. We decompose $\omega$: $\omega =\omega_i^j\sigma_jdx^j$. Due to the additive property of $E_\omega$ and that the action of $C^\infty_c(M)$ commutes with $U_{\omega }$ we only have to analyze an $\omega $ of the form $\sigma_i dx^j$. For a given point $p\in M$ choose the points $$p_0=p,\quad p_1=e^{\frac{1}{n}X}(p),\ldots  ,\quad  p_n= e^{\frac{n}{n}X}(p)$$ 
on the path
$$t\to e^{tX}(p)  ,t\in [0,1].$$
We write the vectorfield $X=X^k\partial_k$. 
We have 
\begin{eqnarray*} 
\lefteqn{e^X_{\nabla+t\omega}}\\
& =&\lim_{n\to \infty } (1+\frac{1}{n}(A(X(p_0)+t\sigma_i X^j(p_0) )) (1+\frac{1}{n}(A(X(p_1))+t\sigma_iX^j(p_1)))\\
&& \cdots (1+\frac{1}{n}(A(X(p_n)+t\sigma_i X^j(p_n)) ,
\end{eqnarray*}
where  $\nabla=d+A$, and therefore 
\begin{eqnarray}
 \lefteqn{\frac{d}{dt}e^X_{\nabla +t\omega}|_{t=0}}
  \label{COOM}
 \\
 &=& \lim_{n\to \infty }  \Big( \frac{1}{n} \sigma_iX^j(p_0)  (1+\frac{1}{n}A(X(p_1)))\cdots (1+\frac{1}{n}A(X(p_n))) \nn\\
 &&+ (1+\frac{1}{n}A(X(p_0))) \frac{1}{n} \sigma_iX^j(p_1)  (1+\frac{1}{n}A(X(p_2)))\cdots (1+\frac{1}{n}A(X(p_n))) \nn\\
 && + \quad\quad\quad\quad\quad\quad\quad\quad\quad\quad\quad \vdots \nn\\
 &&+ (1+\frac{1}{n}A(X(p_0))) )  (1+\frac{1}{n}A(X(p_2)))\cdots (1+\frac{1}{n}A(X(p_{n-1})) \frac{1}{n} \sigma_i X^j(p_n) \Big)
\nn
\end{eqnarray}
We see that before taking the limit $\lim_{n\to \infty}$ this is just the commutator of the sum of the flux operators $\sum_k \frac{1}{n}X^j (p_k) F^{S_k}_i $, where $S_k$ is the plane orthogonal to the $X_j$-axis intersecting $p_k$, and the holonomy operator of the path $$t\to e^{tX}(p)  ,t\in [0,1],$$
 see figure \ref{heltvildtsvedig}.
 
It follows that $E_{\sigma_i dx^j}$ is  a series of flux-operators $F^S_i$ sitting along the path $$t\to e^{tX}(p)  ,t\in [0,1],$$
where the surfaces $S$ are just the planes othogonal to the $x_j$ direction.
But since there is infinitely many of them, they have been weighted with the infinitesimal length, i.e. with a $dx^j$, see figure \ref{heltvildtsvedig}. We can formally write 
$$E_\omega = \int F^S_i dx^i  . $$   

Note that the phenomenon from the holonomy-flux-algebra, that a path $p$ running inside a surface $S$, has zero commutator with the corresponding flux operator is encoded in the quantum-flow-algebra, since the tangent vectors of $p$ will be annihilated by the differential form $dx^j$.

  \subsection{The canonical commutation relations}

We can also make the holonomies infinitesimal, in order see the canonical commutation relations of General Relativity. In the following we scale the translation operators $U_\oo\rightarrow U_{\kappa\oo}$, where $\kappa=8\p G/c^3$. Let us first introduce a coordinate system $(x_1,x_2,x_3)$. We have the vectorfields $\partial_i$, and we consider the operator
$$ \frac{d}{ds} e^{s\partial j}|_{s=0}  .$$ 
We note that if the local connection one form of $\nabla $ is $A_i^j\sigma_jdx^i$ we have 
 $$( \frac{d}{ds} e^{s\partial j}|_{s=0})(\xi (x,\nabla)) = \sigma_i A_j^i(x) \xi (x,\nabla) . $$
We therefore consider the operator $\delta_x \frac{d}{ds} e^{s\partial j}|_{s=0}$, where $\delta_x$ is the delta function located in $x$, as the operator $\sigma_i A_i^j(x)$ located in $x$.  
 
On the other hand consider $\omega =\sigma_ldx^k$. We get 
\begin{eqnarray*}
\lefteqn{[E_{\sigma_ldx^k},\sigma_i A^i_j(x)] (\nabla )}\\
&=& \delta_x  \frac{d}{ds} [E_{\sigma_ldx^k},  e^{s\partial j}  ]|_{s=0}(\nabla)= \kappa \delta_x  \frac{d}{ds} \frac{d}{dt} e^{s\partial j}_{\nabla+t\sigma_ldx^k}|_{t=0}|_{s=0}\\
&=&  \kappa \delta_x  \frac{d}{ds} \frac{d}{dt} (1+s(A_j^i\sigma_i+t\sigma_l \delta_j^ k  ))|_{t=0}|_{s=0}=\kappa \sigma_l \delta_j^k \delta_x \;.
\end{eqnarray*}
If 
$$f_y(x)=\left\{ 
\begin{array}{cl}
1& x=y \\
0& x\not =y
\end{array}\right.$$
we can therefore consider the operator $f_y  E_{\sigma_ldx^k}$ as $\hat{E}_l^k (y)$ since then 
\begin{equation}
 [ \hat{E}_l^k (y) , \sigma_i A^i_j(x)] = \kappa \sigma_l \delta_j^k \delta (x-y),
\label{COOOM}
 \end{equation}
which is the quantized canonical commutation relation of General Relativity formulated in terms as Ashtekar variables.\\

All together these results show that the algebras $\mathbf{QHD}$ and $\mathbf{dQHD}$ are intimately related to canonical quantum gravity since they are simply the algebras from which the infinitesimal operators forming the canonical commutation relations originate.

\section{Representing $\mathbf{HD}$, $\mathbf{QHD}$ and $\mathbf{dQHD}$}

Once the Quantum Holonomy-Diffeomorphism Algebras $\mathbf{QHD}$ and $\mathbf{dQHD}$ are defined the interesting question arises whether representations of these algebras exist. Since we have just demonstrated that these algebras entail the quantized commutation relations of General Relativity, the task of finding representations is equivalent to finding a kinematical Hilbert space. The following section is devoted to this question.


\subsection{Lattice formulation}

\begin{figure}[t]
\begin{center}
\resizebox{!}{2cm}{
 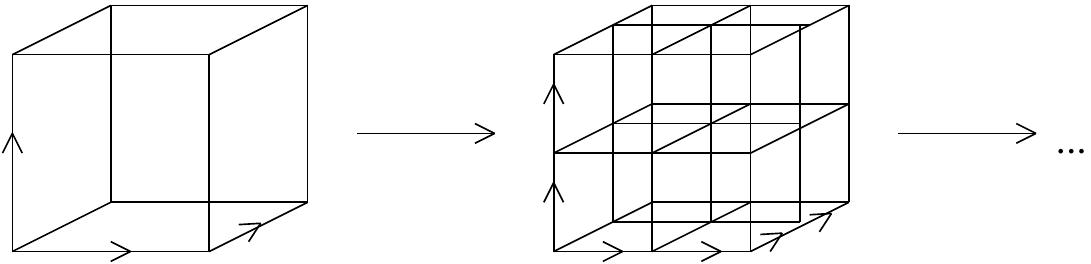}
\end{center}
\label{lenins}
\caption{The lattice approximation is associated to an infinite system of nested cubic lattices.}
\end{figure}

The key tool is to formulate everything in terms of sequences of lattice approximations. This idea and technique was developed in \cite{Aastrup:2012vq} for the holonomy-diffeomorphism algebra $\mathbf{HD}$ and can be extended to $\mathbf{QHD}$ and $\mathbf{dQHD}$ straight forwardly. 

In the lattice formulation an element in $\mathbf{HD}$ is represented by an infinite family of operators acting in increasingly accurate lattice approximations associated to an infinite system $\{\G_n\}$ of nested cubic lattices, see figure 3. This system of lattices corresponds to a coordinate system and thus the lattice formulation of $\mathbf{HD}$ can also be understood as a coordinate dependent formulation.

Here we shall not give details on how the continuum limit of the lattice approximations -- manifolds, operator algebras, Hilbert spaces -- is taken but simply refer the reader to \cite{Aastrup:2012vq}. \\

Thus, we begin with a single finite cubic lattice $\G_n$ with vertices and edges denoted by $\{v_i\}$ and $\{l_j\}$. We assign to each edge $v_i$ a copy of a compact Lie group $G$ 
$$
\nabla(l_j) = g_j\in G
$$
and obtain the space 
$$\ca_{n}= G^{\vert\bf{l}\vert}$$
where $\vert\bf{l}\vert$ is the number of edges in $\G_n$. $\ca_{n}$ is an approximation of the space of connections $\ca$ and the map $\nabla$ should be understood as an approximation of a connection in $\ca$. In the following $G$ will equal $SU(2)$ since this is the Lie group relevant when considering Ashtekar variables. We emphasize, however, that the construction works for other groups too.\\

\begin{figure}[t]
\begin{center}
\resizebox{!}{ 6 cm}{
 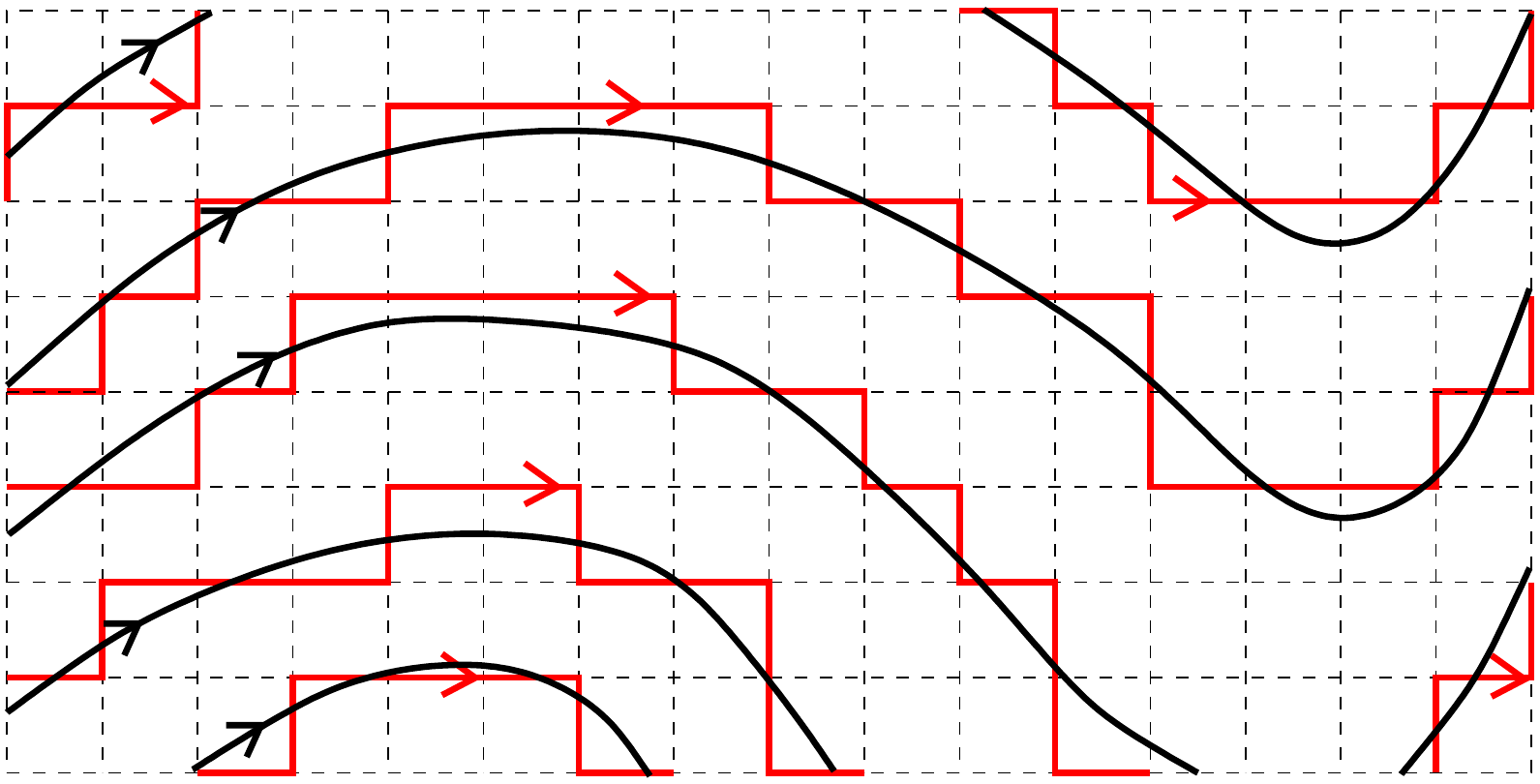}
\end{center}
\caption{\it Sketch of a flow $fF$ and its lattice approximation $\mathbbm{fF}$. Each of the red paths are elements in $\mathbbm{F}$. }
 \label{lennie}
\end{figure}

An element $f e^X$ in $\mathbf{HD}$ is approximated by a finite family of oriented, weighted paths in $\G_n$, denoted by  $\mathbbm{fF}$, see figure \ref{lennie}. 
Here $\mathbbm{F}$ denotes a family $\{ p_i\}$ of paths in $\G_n$, where each $p_i$ is a sequence of adjacent edges 
$$
p_i = \{l_{i_1}, \ldots,l_{i_n}\}\;,
$$
and $\mathbbm{f}$ denote a corresponding set of weights assigned to each edge in $\mathbbm{F}$ (see \cite{Aastrup:2012vq} for details). This lattice approximation of $\mathbf{HD}$ is denoted by $\mathbf{HD}_{n}$.

With the space $\ca_n$ we automatically have the Hilbert space $L^2(\ca_{n})$ via the Haar measure on $G$. We will, however, need more structure in order to construct a representation of $\mathbf{HD}_{n}$  and therefore introduce the Hilbert space
$$
\ch_{n} = L^2(\ca_{n}, \mathbb{C}^2)\times \mathbb{C}^{\vert {\bf l}\vert}
$$
where a representation of an element $\mathbbm{fF}$ in $\mathbf{HD}_{n}$ acts by multiplying the $\mathbb{C}^2$ factor in $\ch_n$ with the parallel transports 
$$
\nabla(p_i) = \nabla(l_{i_1})\cdot \ldots\nabla(l_{i_n})\;,\quad p_i \in \mathbbm{F} \;,
$$
and by acting on the $\mathbb{C}^{\vert {\bf l}\vert}$ factor with $\mathbbm{F}$ as an $\vert {\bf l}\vert \times \vert {\bf l}\vert$-matrix in the sense that each path in $\mathbbm{F}$ shifts lattice points according to its start and end-points. Thus, an example of a representation of an element $\mathbbm{fF}$ in $\mathbf{HD}_{n}$ as an operator in $\ch_n$ could be:
\begin{equation}
\left(
\begin{array}{ccccccccccc}
1 & \ldots & 0 &&&&&&&& \\
\vdots & \ddots & \vdots &&&&&&&&  \\
0 & \ldots & 1 &&&&&&&& \\
&&& \ddots &&&&&&&\\
&&&& 0 & \nabla(p_1) & \nabla(p_2) &&&& \\
&&&& 0 & 0 & \nabla(p_3) &&&& \\
&&&& 0 & 0 & 0 &&&& \\
&&&&&&& \ddots &&& \\
&&&&&&&& 1 & \ldots & 0 \\
&&&&&&&& \vdots & \ddots & \vdots \\
&&&&&&&& 0 & \ldots & 1
\end{array}
\right) \nn
\end{equation}
where $\mathbbm{F}$ involves\footnote{In order to ease the notation we have not assigned weights to the paths in this example. If there had been weights these would have appeared as numbers multiplied the $\nabla(p_i)$'s.} the three paths $\{p_1,p_2,p_3\}$.\\

The additional $\mathbb{C}^{\vert {\bf l}\vert}$ factor in $\ch_n$ corresponds in the continuum limit to a tensor product with $L^2(M)$. Thus, in this limit the inner product in $\ch_n$ involves an integral over $M$.

\subsection{Approximation of the translation operator $E_\oo$}

In \cite{Aastrup:2012vq} we only presented lattice approximations of the $\mathbf{HD}$ algebra.  In order to obtain lattice approximations of $\mathbf{QHD}$ and $\mathbf{dQHD}$ we also need to define the operators $E_\oo$ and $U_\oo$ on the lattice.

To do so we first choose a left and right invariant metric $\langle \cdot,\cdot\rangle$ on G. Let $\mathfrak{g}_i$ be the Lie algebra of the $i$'th copy of $G$ and choose an orthonormal basis $\{e_i^a\}$ for $\mathfrak{g}_i$ with respect to $\langle \cdot,\cdot\rangle$. Here the index $i$ labels the different copies of $G$ while $a$ is an index for the Lie algebra of $G$.  We also denote by $\{{\bf e}_i^a\}$ the corresponding left translated vector fields and by $\cl_{{\bf e}_i^a}$ the derivation with respect to the trivialization given by $\{e_i^a\}$. 
There is then a natural candidate for a lattice approximation of $E_\oo$ given by 
$$
 E_{\oo,n}=  2^{-n}\kappa \sum_{i,a} \oo_i^a   \cl_{{\bf e}_i^a} \otimes 1_{\vert \bf{l}\vert \times \vert \bf{l}\vert}
$$
where $\oo_i^a$ here denotes the value of $\oo$ evaluated at the start-point of the edge $l_i$ and where the sum runs over all edges in the lattice and over all $\mathfrak{g}$-indices. One can easily check that the commutator between  $E_{\oo,n}$ and a parallel transport $\nabla(p_i)$ has the same structure as the commutator between the operator $E_{\oo}$ and an element in $\mathbf{HD}$, see equation (\ref{COOM}). Likewise, in \cite{AGNP1} we have shown that the commutator between the vector-field $\cl_{{\bf e}_i^a} $ and $\nabla(p_i)$ reproduces the structure of the Poisson bracket of general relativity formulated in terms of Ashtekar variables, the bracket, which we also found in (\ref{COOOM}). 

For the lattice approximation of the translation operator $U_\oo$ one can expect to simply exponentiate the operator $E_{\oo,n}$.

The convergence conditions for the operator $E_{\oo,n}$ and its exponentiation are identical to convergence conditions for operators in $\mathbf{HD}$ defined in \cite{Aastrup:2012vq} and shall not be given here.\\


\subsection{Conditions on lattice states}

We are now ready to write down necessary conditions for states on $\mathbf{HD}$ and $\mathbf{dQHD}$. Consider first a lattice approximation $\G_n$ and a vector state $\xi_n$ in $\ch_n$
$$
\xi_n = \P_{i=1}^{\vert {\bf l}\vert} \xi_{(n,i)}\otimes \eta_n
$$
where $\xi_{(n,i)}$ is  function on the $i$'th copy of $G$ in $\ca_n$, and $\eta_n$ is a $\mathbb{C}^2$-valued function on the set of vertices in $\G_n$ (When $n\to \infty$ the function $\eta_n$ corresponds to a $\mathbb{C}^2$-valued function $\eta$ on $M$).  Denote by $\xi$ the limit state given by an infinite sequence $\{\xi_n\}_{n\in\mathbb{N}_+}$.
The necessary conditions for $\xi$ to be a state on $\mathbf{HD}$ and $\mathbf{dQHD}$ are the following
\begin{eqnarray}
1.&&\langle \xi_{(n,i)} \vert \nabla(l_i) \vert \xi_{(n,i)}\rangle = \mathrm{1} + \co(dx)
\nn\\
2.&&\langle \xi_{(n,i)} \vert \cl_{{\bf e}_{i_1}}\cl_{{\bf e}_{i_2}}\cdot\ldots\cdot \cl_{{\bf e}_{i_m}}  \vert \xi_{(n,i)}\rangle = \co(dx^{2m})\;,\quad \forall m\;, i_n\in\{1,2,3\}
\nn
\end{eqnarray}
where $dx=2^{-n}$.

Condition $1.$ ensures the the expectation value of a parallel transport $\nabla(p)$ will converge in the limit $n\rightarrow\infty$. This condition is easily satisfied and we may therefore use the GNS construction to obtain the Hilbert space $\ch_{\xi_n}$ and its limit $\ch_\xi$. Furthermore, we may write
\begin{equation}
\langle \xi_{(n,i)} \vert \nabla(l_i) \vert \xi_{(n,i)}\rangle = \mathrm{1} + dx (A_{(n,i)} + t B_{(n,i)})
\label{Ash}
\end{equation}
where $A_{(n,i)}$ is the self-adjoint component, $B_{(n,i)}$ the remainder and $t$  a formal parameter. If we let $t$ play the role of a quantization parameter we see that the states on $\mathbf{HD}$ constructed this way will always be semi-classical with $A$ being the classical point and $B$ the quantum correction.\\

Condition $2.$ ensures that the expectation value of the translation operators $E_{\oo,n}$ and all polynomials thereof will converge in the limit $n\rightarrow\infty$. However, we strongly expect that this condition can never be satisfied simultaneous with condition $1$, since condition 1. ensures, that the state is peaked around the identity of the group, whereas the second condition more or less gives that the state is constant. 
 \\

To sum up, we make the following conclusions:
\begin{enumerate}
\item
There is a state $\xi$ on $\mathbf{HD}$ and by the GNS construction a Hilbert space $\ch_\xi$. Also, the state $\xi$ will always be a semi-classical state.
\item
There are, within the framework of lattice approximations, no states on $\mathbf{dQHD}$. The infinitesimal translation operators $E_\oo$ are not well defined operators in $\ch_\xi$.
\end{enumerate}

Whether the negative result regarding states on $\mathbf{dQHD}$ and $\mathbf{QHD}$ applies only to states constructed via lattice approximations or whether it is generic -- that no states exist on $\mathbf{QHD}$ and $\mathbf{dQHD}$ -- is not clear. We are inclined to believe it is generic, but further analysis is needed in order to give a definite answer.\\


\section{Unbounded operators via the state $\xi$}

In order to proceed we now ask the following question: {\it what physically significant unbounded operators can we define via a GNS-construction with state $\xi$?} It may be that there exist a sub-algebra in $\mathbf{dQHD}$ or some related algebra, including and larger than $\mathbf{HD}$, on which we can use $\xi$ to define a GNS-representation.

It turns out that such operators do appear to exist and that exactly these operators are very significant from a physical perspective. The following section deals with this question.


\subsection{A Dirac type operator}

To find unbounded operators acting in $\ch_\xi$ we shall define a Dirac-type operator and analyze its commutators. 
We therefore return to the lattice approximation and the operator $E_{\oo,n}$. We first add an additional factor to the Hilbert space $\ch_n$
$$
\ch'_n  = L^2(\ca_{n}, Cl(T^*\ca_n)\times \mathbb{C}^2)\times \mathbb{C}^{\vert {\bf l}\vert}
$$
where $Cl(T^*\ca_n)$ is the Cliford algebra over the co-tangent space over $\ca_n$. This enables us to turn $E_{\oo,n}$ into a Dirac-type operator over the space $\ca_n$ by substituting $\oo$ in $E_{\oo,n}$ with an element of the Clifford algebra:
$$
E_{\oo,n}  \stackrel{\oo\rightarrow{\bf e}}{\longrightarrow} D_n = 2^{-n}\kappa \sum_{i,a}  {\bf e}_i^a \cdot \cl_{{\bf e}_i^a} \otimes 1_{\vert \bf{l}\vert \times \vert \bf{l}\vert}
$$
where we by ${\bf e}_i^a$ also denote the element of $Cl(T^*\ca_n)$ that corresponds to the left-translated vector field ${\bf e}_i^a$ and where '$\cdot$' denotes Clifford multiplication. The operator $D_n$ is essentially the Dirac type operator, which we studied in the papers \cite{Aastrup:2012jj}-\cite{AGNP1} . 

Here, however,  we shall require a different presentation of the Clifford algebra, which means that we shall assume that the Clifford elements fullfill the following relation
\begin{equation}
\{ {\bf e}_i^a, {\bf e}_j^b   \} =  2^{n}  \d^{ab}  \d_{ij}\;.
\label{Cliff}
\end{equation}
This means that the anti-commutator, in the large $n$ limit, will approach a one-dimensional delta function. Of course, if ${\bf f}_i^a$ denotes the standard generators, we get the above generators via $ {\bf e}_i^a=2^{\frac{n}{2}}{\bf f}_i^a$.

Next, define the operation $\d_n$, which consist of taking the commutator with $D_n$
$$
\d_n a_n := [a_n, D_n]\;,\quad a_n\in \cb(\ch'_n)
$$
where the commutator is graded. One can check that for $a_n\in\mathbf{HD}_{n}$ $\d_n a_n$ will be a sum over insertions in $a_n$ of the form
$$
2^{-n} \kappa \sigma^a  {\bf e}^a_j  
$$
where $j$ now labels the point of insertion in $a_n$. We now consider the second commutator
$$
\d_n^2 a_n =[D_n,[D_n,a_n]]
$$
and find that this expression will be a sum of insertions of the form
\begin{equation}
2^{-n} \kappa \sigma^a \cl_{{\bf e}_j^a}  \;.
\label{august}
\end{equation}
From this we can conclude that $\d_n^2 a_n$ descends, in the limit $n\rightarrow\infty$, to a well defined operator in $\ch_\xi$, which we denote $\d^2 a$ where $a$ is an element in $\mathbf{HD}$. 

The reason why $\d_n^2 a_n$ descents to a well defined operator using $\xi$ is that the factor $2^{-n}$ is exactly the factor required for a sequence of insertions in $a_n$ to converge in the $n\rightarrow\infty$ limit. This explains why we needed to introduce a scaled version of the Clifford algebra in (\ref{Cliff}).

We shall here not concern ourselves with the question of what role the Dirac type operator $D_n$ plays in the $n\rightarrow\infty$ limit. In fact, one may completely ignore the Dirac type operator construction and just use the form of the insertions (\ref{august}) to define the operator $\d_n^2 a_n$ without any reference to the previous analysis. It is, however, our belief - also in the light of the material of the next section - that the Dirac type operator is of significance.  \\

The operators $\d^2_n a_n$ are very interesting seen from a physical perspective. The quantity
$$
 \kappa \sigma^a g_j \cl_{{\bf e}_j^a}   \;,
$$
which arises in $\d_n^2 a_n$, is  the type of operator that we found in \cite{Aastrup:2012jj}-\cite{AGNP1}  to entail the spatial 3-dimensional Dirac operator in a combined continuum and semi-classical limit. To obtain this result one must put conditions on the expectation value of the vector-fields. Here we shall not give the details of how this works, but simply conclude that the unbounded operators, that we have just identified, are the type of operators, which entail spatial Dirac operators in a semi-classical limit.

\section{On the dynamics of GR}

Let us now proceed by analyzing higher orders of $\d_n $. In this subsection we shall only be concerned with the general structure and leave the details to be worked out in a later publication. 

Consider first a one-form\footnote{In the terminology of noncommutative geometry, $B_n$ is a connection one-form over the (approximate) space $\ca_n$ of connections. }
$$
B_n = a_0 \d_n a_1 \;,\quad a_0,a_1\in\mathbf{HD}_n\;,
$$
where we shall assume that $a_0$ and $a_1$ connect adjacent vertices in $\G_n$. Thus, $a_0$ and $a_1$ corresponds to elements in $ \mathbf{HD}$, which are infinitesimal with respect to the manifold $M$. Also, consider the operator
$$
\d_n B_n 
$$
where we find that the lowest order entries (with respect to factors of $2^{- n}$) in $\d_n{B_n} $ are of the form
\begin{equation}
2^{-n} \kappa\cl_{{\bf e}_k^b} g_i \sigma^a  g_j \{ {\bf e}^a_j, {\bf e}^b_k \} =  \kappa\cl_{{\bf e}_j^a} g_i  \sigma^a   g_j  \;.
\nn
\end{equation}
Let us now consider the square of $\d_n {B} $
$$
 \d_n{B_n} (\d_n{B_n})^*\;,
$$
which represents a curvature operator over the space $\ca_n$.
We find that the only terms of significance are of the form
$$
 \kappa^2 \cl_{{\bf e}_j^a} g_i g_j  \sigma^a   \sigma^b  g_k g_l   \cl_{{\bf e}_k^b} \;.
$$
Here it is important to notice that the two derivatives will be associated to adjacent edges. If we take the expectation value of this quantity (again: we ignore the $2^{\pm n}$ factors)
\begin{equation}
\lim_{n\rightarrow\infty}\lim_{\kappa\rightarrow 0}\langle \xi_n \vert  \d_n{B_n} (\d_n{B_n})^*  \vert \xi_n \rangle \sim \int_M d^3x  \Big(\eta (x) | E_a^m(x) E^n_b(x) F_{mn}^c(x) |\eta (x) \Big)
\label{dynamics}
\end{equation}
where we assumed a condition on the expectation value of the vector field
$$
\lim_{\kappa\rightarrow 0}\langle \xi_{(n,i)} \vert \kappa\cl_{{\bf e}_{i}^a}\vert \xi_{(n,i)}\rangle \sim \mathrm{i}E^{m}_a (x)
$$
where $x$ is the endpoint of the $i$'th edge in $\G_n$ and where $m$ is its spatial orientation, $m\in\{x,y,z\}$ and where $E^m_a(x)$ is a densitized triad field. Furthermore, $F$ is the curvature of the connection $A$ and we assume that the four different group elements $g_i,g_j,g_k, g_l$ are associated to a sequence of edges $\{l_i,l_j,l_k,l_l\}$ which form a closed loop. 

Of course if we wanted an expression like  
$$\int_M d^3x  \Big( E_a^m(x) E^n_b(x) F_{mn}^c(x) \e^{ab}_{\;\;\;c} \Big),$$
one would get this by taking a partial trace of  $\d_n{B_n} (\d_n{B_n})^*$, i.e. something like
$$ \lim_{n\rightarrow\infty}\lim_{\kappa\rightarrow 0}\langle \xi_n \vert Tr_{\mathcal{B} (\mathbb{C}^2)} (\d_n{B_n} (\d_n{B_n})^* ) \vert \xi_n \rangle ,$$
by letting the trace act diagonally in $\mathbb{C}^2$ and letting the spinor-field $\eta$ be constant equal to $(1,0)$. 

The rhs of equation (\ref{dynamics}) has the structure of the Hamiltonian of General Relativity formulated in terms of Ashtekar variables. Thus, this computation suggest that the dynamics of General Relativity emerges from the positive definite operator $  \d_n{B_n} (\d_n{B_n})^*$ representing the curvature on the space of connections.


\section{Conclusion}

We have introduced the Quantum Holonomy-Diffeomorphism Algebra and shown that it encodes the canonical commutation relations of quantum gravity formulated in terms of Ashtekar variables. We have argued that a state built over an infinite sequence of lattice approximations exist on the Holonomy-Diffeomorphism part of the algebra, and that this state cannot be extended to the full algebra. Furthermore, we have argued that a certain class of unbounded operators, that act in the GNS construction of this state, does exist. These unbounded operators are exactly the type of operators, which we have previously analyzed and found to entail a classical 3-dimensional Dirac operator in a semi-classical limit. We find these states via a construction that involves a Dirac type operator over the space of connections. Finally, a simple computation shows that the structure of the Hamiltonian of General Relativity emerges from a curvature operator constructed via this Dirac type operator.

\vspace{0.5cm}
\noindent{\bf Acknowledgements}\\
We are thankful to Mario Paschke for useful comments and suggestions.

\end{document}

%% file: ogsaasvedig.pdf_t.tex
\begin{picture}(0,0)%
\includegraphics{ogsaasvedig.pdf}%
\end{picture}%
\setlength{\unitlength}{4144sp}%
\begingroup\makeatletter\ifx\SetFigFont\undefined%
\gdef\SetFigFont#1#2#3#4#5{%
  \reset@font\fontsize{#1}{#2pt}%
  \fontfamily{#3}\fontseries{#4}\fontshape{#5}%
  \selectfont}%
\fi\endgroup%
\begin{picture}(7029,4859)(1094,-5708)
\end{picture}%

%% file: svedigtegning1.pdf_t.tex
\begin{picture}(0,0)%
\includegraphics{svedigtegning1.pdf}%
\end{picture}%
\setlength{\unitlength}{4144sp}%
\begingroup\makeatletter\ifx\SetFigFont\undefined%
\gdef\SetFigFont#1#2#3#4#5{%
  \reset@font\fontsize{#1}{#2pt}%
  \fontfamily{#3}\fontseries{#4}\fontshape{#5}%
  \selectfont}%
\fi\endgroup%
\begin{picture}(9839,3644)(999,-3683)
\end{picture}%

%% file: AAAcubes.pdf_t
\begin{picture}(0,0)%
\includegraphics{AAAcubes.pdf}%
\end{picture}%
\setlength{\unitlength}{4144sp}%
\begingroup\makeatletter\ifx\SetFigFont\undefined%
\gdef\SetFigFont#1#2#3#4#5{%
  \reset@font\fontsize{#1}{#2pt}%
  \fontfamily{#3}\fontseries{#4}\fontshape{#5}%
  \selectfont}%
\fi\endgroup%
\begin{picture}(3894,1239)(3544,-5518)
\end{picture}%

%% file: flowapproxred.pdf_t.tex
\begin{picture}(0,0)%
\includegraphics{flowapproxred.pdf}%
\end{picture}%
\setlength{\unitlength}{4144sp}%
\begingroup\makeatletter\ifx\SetFigFont\undefined%
\gdef\SetFigFont#1#2#3#4#5{%
  \reset@font\fontsize{#1}{#2pt}%
  \fontfamily{#3}\fontseries{#4}\fontshape{#5}%
  \selectfont}%
\fi\endgroup%
\begin{picture}(7266,3688)(868,-3709)
\end{picture}%